# Physical Tests for Random Numbers in Simulations

I. Vattulainen,[1,2] T. Ala–Nissila,[1,2] and K. Kankaala[2,3]

[1]Research Institute for Theoretical Physics
P.O. Box 9 (Siltavuorenpenger 20 C)
FIN–00014 University of Helsinki
Finland

[2]Department of Electrical Engineering
Tampere University of Technology
P.O. Box 692, FIN–33101 Tampere
Finland

[3]Center for Scientific Computing
P.O. Box 405, FIN–02101 Espoo
Finland

May 30, 1994

## Abstract

We propose three physical tests to measure correlations in random numbers used in Monte Carlo simulations. The first test uses autocorrelation times of certain physical quantities when the Ising model is simulated with the Wolff algorithm. The second test is based on random walks, and the third on blocks of $n$ successive numbers. We apply the tests to show that recent errors in high precision simulations using generalized feedback shift register algorithms are due to short range correlations in random number sequences. We also determine the length of these correlations.

The Monte Carlo (MC) simulation method is a standard technique in physical sciences [1]. The key ingredient in its successful application lies in the quality of random numbers used, which are usually produced by deterministic pseudorandom number (PRN) generator algorithms. Several tests for PRN generators have been suggested [2, 3, 4] and conducted [3, 4, 5, 6] to test PRN generators, but none of them can prove that a given generator is reliable in *all* applications. Sometimes the inevitable correlations in their output can lead to erroneous results, as recently demonstrated in high–precision MC simulations for some commonly used generators combined with special simulation algorithms [7, 8, 9, 10, 11].

In this letter, our purpose is to introduce physical tests which allow more precise characterizations of correlations which may cause problems in simulations. These tests are then used to demonstrate that for some generators, there exist nontrivial *local* correlations which are present in rather short subsequences of random numbers, and cannot always be detected by conventional test methods [2, 3, 6]. We start by generalizing the Ising model simulations of Refs. [7, 10, 11] to show that the local correlations lead to *deviations in the cluster formation process* of the Wolff algorithm [12] which explains the errors observed in Refs. [7, 10, 11]. As a test of random numbers, we show that *integrated autocorrelation times* [13] of certain physical quantities are particularly sensitive measures of local correlations. We then introduce *random walk* and *n–block* tests, and show how these tests can be used to precisely measure the length of correlations in PRN sequences. In particular, for the generalized feedback shift register (GFSR) generators we show that the length of correlations is very close to the longer lag parameter of the generator.

The PRN generators tested in this work include GFSR algorithms [14], which are of the form $x_n = x_{n-p} \oplus x_{n-q}$, where $\oplus$ is the bitwise exclusive OR operator. They are denoted by R$p$; recommended values for $p$ and $q$ ($p > q$) can be found $e.g.$ in Refs. [15]. Other generators include a linear congruential generator $x_n = (16807 \times x_{n-1}) \bmod (2^{31} - 1)$ [16] known as GGL (CONG in Ref.[7]), RAN3 [17], which is a lagged Fibonacci generator $x_n = (x_{n-55} - x_{n-24}) \bmod 2^{31}$, and a combination generator RANMAR [18]. The GFSR generators were initialized with 32–bit integers produced by GGL. Other initialization methods including the one in Ref. [19] were also checked but the results were unaffected.

In Refs. [7, 10, 11] problems with GFSR generators arose when the two–dimensional Ising model was simulated on a square lattice with the Wolff algorithm [12]. We carried out analogous simulations for an Ising model of linear size $L = 16$ at $K_c = \frac{1}{2}\ln(1 + \sqrt{2})$, for a variety of generators listed in Table 1. Our implementation of the single cluster search algorithm follows Ref. [20]. We calculated the energy $E$, the magnetic susceptibility $\hat{\chi}$, and the (normalized) size of the flipped clusters $c$. We then calculated the corresponding *integrated autocorrelation times* $\tau_E, \tau_{\hat{\chi}}$, and $\tau_c$, by first calculating the autocorrelation functions

$$C(t) = \frac{\langle A(t_0)A(t_0 + t)\rangle - \langle A(t_0)\rangle^2}{\langle A(t_0)^2\rangle - \langle A(t_0)\rangle^2}, \tag{1}$$

and then following the procedure in Ref. [13].

A summary of the results in Table 1 shows that based on this test, the generators can be classified into two categories. For the energy $\langle E \rangle$, for example, deviations from the exact result of $\langle E \rangle = 1.45312$ [21] for R31, R250, R521, and RAN3 are much larger than $3\sigma$ where $\sigma$ is the standard deviation [13]. In particular, $\langle c \rangle$ reveals that in the erroneous cases the *average flipped cluster size* is biased. Most striking, however, is the behavior of the integrated autocorrelation times. For generators, which show no significant deviations in $\langle E \rangle$, $\langle \hat{\chi} \rangle$, or $\langle c \rangle$, results for the $\tau$'s agree well with each other. However, for R31 and R250 the integrated autocorrelation times show errors of about 8%. We thus propose these quantities as particularly sensitive measures of correlations in PRN sequences.

To compare with Refs. [7, 10] we also used the autocorrelation time test to study the decimation of the output of R250, *i.e.* we took every $k$th number of the PRN sequence. For $k = \{3, 5, 6, 7, 9, 10, 11, 12, 24, 48\}$ the correlations vanish in agreement with $k = 5$ in Ref. [7] and $k = 3, 5$ in Ref. [10]. On the other hand, for $k = 2^m$ with $m = \{0, 1, 2, 3, 4, 5, 6, 7, 8\}$ the sequences fail. These findings agree with the theoretical result of Golomb [22], who showed that the decimation of a maximum−length GFSR sequence by powers of two results in statistically equivalent sequences.

The errors in the average cluster sizes for some of the GFSR generators suggests that there must be correlations present within the $\mathcal{O}(L^2)$ successive PRN's which are used in the single cluster formation of the Wolff algorithm. To quantify the range of correlations we propose the following *random walk* and *n−block* tests.

In the *random walk test* [23], we consider random walks on a plane which is divided into four equal blocks, each of which has an equal probability to contain the random walker after a walk of length $n$. The test is performed $N$ times, and the number of occurrences in each of the four blocks is compared to the expected value of $N/4$, using a $\chi^2$ test with three degrees of freedom. The generator fails if the $\chi^2$ value exceeds 7.815 in at least two out of three independent runs. This should occur with a probability of only about 3/400.

Results for a group of generators with this test are presented in Table 2 with $n = 1000$. They are in agreement with the autocorrelation time test. No correlations for either GGL or RANMAR were observed. R250 and R521 pass the test with $k = 3$, but fail with values $k = \{1, 2, 2^6\}$, whereas R1279 passes with all $k$ tested. The failure of RAN3 with $k = 1$ is consistent with previous test results [4, 6]. It is notable, that all the failures in this test were very clear, since even the smallest $\chi^2$ values exceeded 40. However, RAN3 passed the test when every second or third number was used.

The main difference between the failing generators R250 and R521 (with $k = 1$) and the successful ones R1279 and R4423 lies in the lag parameter $p$ which is less than $n$ for the former and larger than $n$ for the latter. Thus, it is plausible that the range of correlations depends on $p$ and $q$. We studied this for $p$ with the random walk test by locating the approximate value $n_c$, above which the generators fail. The test was performed for R31, R250, R521 and R1279 with $N = 10^6$ samples. The results for $n_c$ were $32 \pm 1$, $280 \pm 5$, $590 \pm 5$, and $1515 \pm 5$, respectively, where the error estimate is the largest distance between samples close to $n_c$. An example of the $\chi^2$ values is given in Fig. 1. These results demonstrate that the correlations are *local*, in the sense that they have a range very close to $p$.

In order to further quantify the nature of correlations we present the $n-block$ test [24]. In it we take a sequence $\{x_1, x_2, \ldots, x_n\}$ of uniformly distributed random numbers $0 \leq x_i < 1$, whose average $\bar{x}$ is calculated. If $\bar{x} \geq 1/2$, we choose $y_i = 1$; otherwise $y_i = 0$. This is repeated $N$ times. We then perform a $\chi^2$ test on variables $y_i$ with one degree of freedom. Each test is repeated three times, and the generator fails the test with fixed $n$ if at least two out of three $\chi^2$ values exceed 3.841, which should occur with a probability of about $3/400$.

In cases of GGL, RANMAR and RAN3, we observed no correlations up to $n = 10^4$ for $N = 10^6$. For R31, R250 and R521 we performed an iterative study by varying $n$. When $N = 10^6$ samples were taken, the resulting correlation lengths $n_c$ were $32 \pm 1$, $267 \pm 5$, and $555 \pm 5$, respectively. With better statistics $N = 10^8$, we observed no change for R31, whereas the estimate for R521 reduced to $525 \pm 1$, and that of R250 to $251 \pm 1$. The latter value was confirmed with $N = 10^9$ also. Typical values of $\chi^2$ for R250 are shown in Fig. 2, where a sharp onset of correlations at $n_c$ is visible.

The results of the random walk and $n-block$ tests together show that for the GFSR generators, the origin of the errors in the simulations presented here and in Refs. [7, 8, 9, 10, 11] must be the appearance of local correlations in the probability distribution. Moreover, our tests show that the length of correlation lies very close to the value of the longer lag parameter $p$. These results are in accordance with a remark in Ref. [25], and quantify the observations of Ref. [9] where it was estimated that "triple" correlations exist about 400 numbers apart for R250. It is important to realize that in our Ising simulations it is the single cluster search [20] in the Wolff algorithm, where $\mathcal{O}(L^2)$ *successive* PRN's are used in cluster formation that makes the system especially sensitive to correlations; if GFSR generators with $p \gg L^2$ ($q \approx p/2$) are used, the results improve considerably. The same can also be achieved with a judicious decimation (*e.g.* $k = 3$) of the sequences.

In conclusion, we have presented three simple physical tests for detecting correlations in PRN sequences. We have demonstrated the quality of these tests by being able to unravel the range and nature of correlations in GFSR generators that have recently been shown to produce erroneous results in Monte Carlo simulations. In particular, we have shown that the origin of these errors lies in local correlations present in the cluster formation process of the Wolff algorithm. It is remarkable that these correlations cannot be seen in either careful statistical [6] or bit level tests [4] but only with the present test methods, and other special simulation algorithms [8, 9]. We believe that together with statistical and bit level tests the physical tests presented here form a rather complete test bench which can be used to develop better generators for demanding applications.

Finally, we would also like to note that we have preliminary results for applying the tests presented here for GFSR generators with four "taps" of the form $x_n = x_{n-9689} \oplus x_{n-471} \oplus x_{n-314} \oplus x_{n-157}$ [26], which is basically a 3–decimation ($k = 3$) of the generator $x_n = x_{n-9689} \oplus x_{n-471}$ [27]. For these generators correlations appear to be very weak as expected, but for smaller lags errors can again be found [28].

We would like to thank R. Ziff for correspondence and providing us with his results prior to publication. K.K. thanks Paul Coddington for useful discussions. Tampere University of Technology and the University of Helsinki have provided ample computing resources. This work has been partly supported by the Academy of Finland.

Internet addresses: `Ilpo.Vattulainen@csc.fi, ala@phcu.helsinki.fi, Kari.Kankaala@csc.fi`

# Table captions

TABLE 1. Results of simulations for the Ising model at criticality with the Wolff algorithm. The number of samples is $N = 10^7$ and $k$ is the decimation parameter. The errors shown correspond to $\sigma$ [13]. The most erroneous results are in boldface. See text for details.

TABLE 2. Results of the random walk test with $N = 10^6$ samples. See text for details.

# Figure captions

FIGURE 1. The $\chi^2$ values for R31 (inner figure) and R250 in the random walk test as a function of walk length $n$. Three independent runs in both cases are denoted by different symbols. The horizontal lines denote $\chi^2 = 7.815$.

FIGURE 2. The $\chi^2$ values for R250 in the $n$-block test. Curves with circles and squares correspond to $N = 10^8$ and $N = 10^9$ samples, respectively. In both cases three independent runs have been performed. The horizontal line denotes $\chi^2 = 3.841$.

| Generator | $k$ | $q$ | $\langle E \rangle$ | $\langle \chi \rangle$ | $\langle c \rangle$ | $\tau_E$ | $\tau_{\chi}$ | $\tau_c$ |
|---|---|---|---|---|---|---|---|---|
| R31 | 1 | 3 | **1.46774(7)** | **0.564(2)** | **0.5664(3)** | **1.233(4)** | **1.058(3)** | **0.507(2)** |
| R250 | 1 | 103 | **1.45509(7)** | 0.548(2) | **0.5474(2)** | **1.333(4)** | **1.143(4)** | **0.589(4)** |
| | 3 | 103 | 1.45302(7) | 0.545(2) | 0.5452(2) | 1.446(6) | 1.226(5) | 0.628(5) |
| R521 | 1 | 168 | **1.45379(7)** | 0.546(2) | **0.5461(2)** | **1.384(5)** | **1.182(5)** | **0.604(4)** |
| R1279 | 1 | 418 | 1.45312(7) | 0.545(2) | 0.5454(2) | 1.426(5) | 1.215(4) | 0.622(3) |
| R2281 | 1 | 1029 | 1.45311(7) | 0.545(2) | 0.5456(2) | 1.439(5) | 1.226(5) | 0.627(5) |
| R4423 | 1 | 2098 | 1.45303(7) | 0.545(2) | 0.5454(2) | 1.441(5) | 1.226(5) | 0.624(4) |
| R9689 | 1 | 4187 | 1.45313(7) | 0.546(2) | 0.5455(2) | 1.444(5) | 1.229(5) | 0.625(4) |
| R19937 | 1 | 9842 | 1.45294(7) | 0.545(2) | 0.5452(2) | 1.434(5) | 1.220(5) | 0.624(4) |
| R44497 | 1 | 21034 | 1.45292(7) | 0.545(2) | 0.5452(2) | 1.434(5) | 1.219(5) | 0.622(2) |
| RAN3 | | | **1.45254(7)** | 0.545(2) | 0.5446(2) | 1.447(5) | 1.231(5) | 0.630(3) |
| GGL | | | 1.45309(7) | 0.545(2) | 0.5454(2) | 1.436(5) | 1.221(5) | 0.622(4) |
| RANMAR | | | 1.45303(7) | 0.545(2) | 0.5452(2) | 1.443(5) | 1.227(5) | 0.624(4) |

Table 1:

| Generator | $k$ | $q$ | Result |
|---|---|---|---|
| R31 | 1 | 3 | FAIL |
| R250 | 1,2,64 | 103 | FAIL |
| R521 | 1,2,64 | 168 | FAIL |
| RAN3 | 1 | | FAIL |
| R250 | 3 | 103 | PASS |
| R521 | 3 | 168 | PASS |
| R1279 | 1,2,3,64 | 418 | PASS |
| R4423 | 1 | 2098 | PASS |
| RAN3 | 2,3 | | PASS |
| GGL | | | PASS |
| RANMAR | | | PASS |

Table 2:

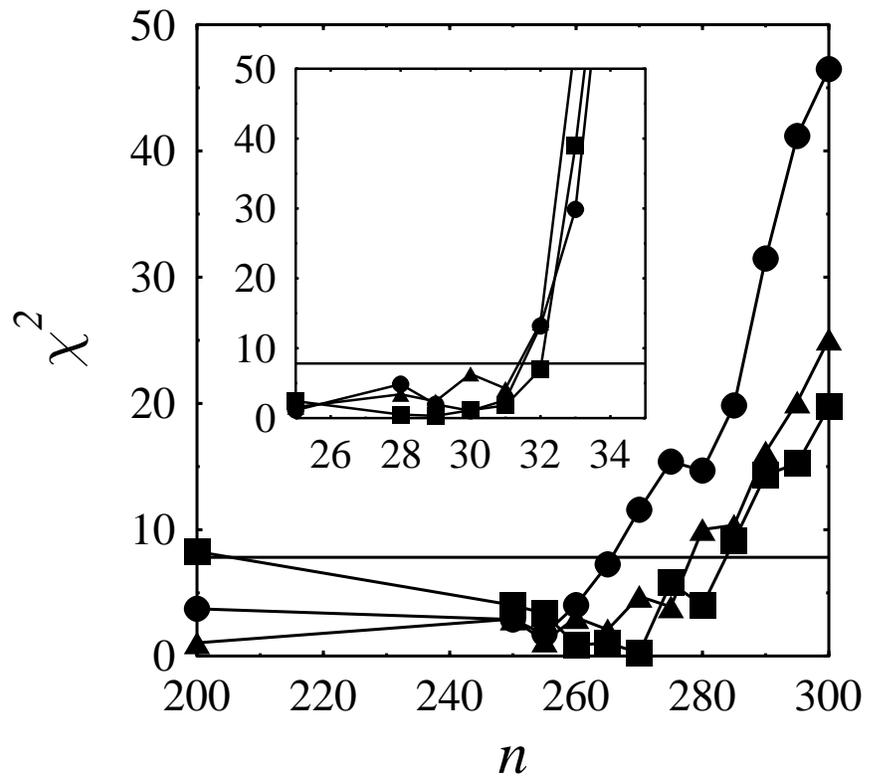

Figure 1:

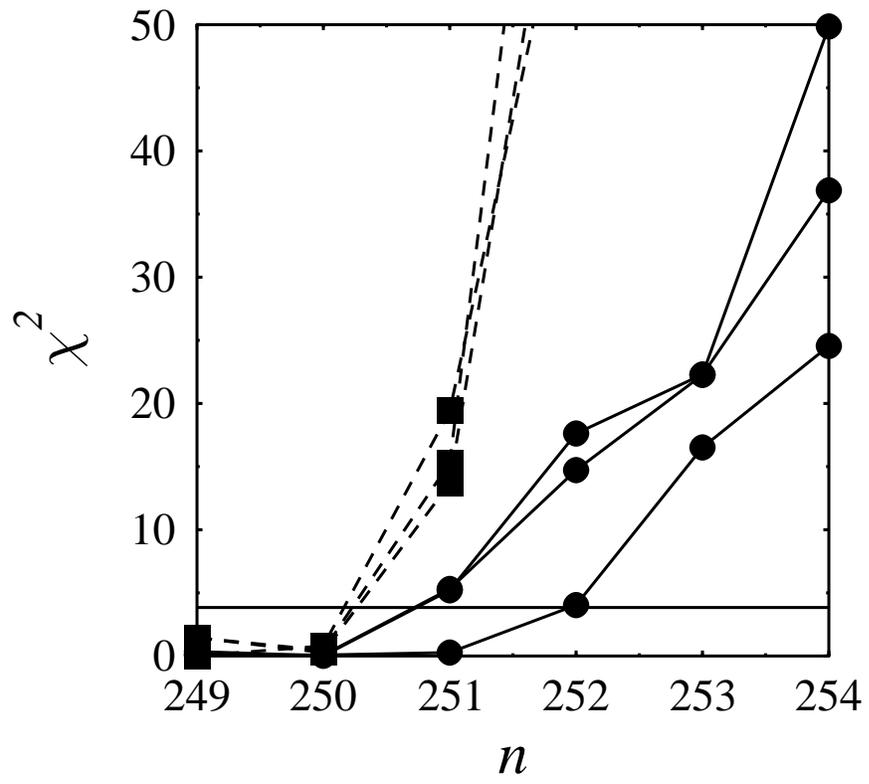

Figure 2: